\documentclass[showpacs, pra,onecolumn,preprintnumbers ,amsmath, amssymb, superscriptaddress, aps]{revtex4-2}
\usepackage{color}
\usepackage{amsmath,amssymb}
\usepackage{pifont}
\usepackage{amssymb}  
\usepackage{bbold}
\usepackage{float}
\usepackage{subfloat}
\usepackage[squaren]{SIunits}
\usepackage[caption=false]{subfig}
\usepackage{tikz}
\usepackage{makecell}
\usepackage{subfig}
\usepackage{pifont}   
\usepackage{graphicx} 
\graphicspath{{Figures/}}
\usepackage{dcolumn}  
\usepackage{bm}       
\usepackage{multirow} 
\usepackage{placeins}
\usepackage[colorlinks]{hyperref}
\usepackage{mathtools}
\usepackage{appendix}

\begin{document}
	
	\title{Spin-dependent transmission in curved graphene superlattice}
	\date{\today}
	\author{Jaouad El-hassouny}
	\affiliation{Laboratory of Nanostructures and Advanced Materials, Mechanics and Thermofluids, Faculty of Sciences and Techniques, Hassan II University, Mohammedia, Morocco}
	
	\author{Ahmed Jellal}
	\email{a.jellal@ucd.ac.ma}
	\affiliation{Laboratory of Theoretical Physics, Faculty of Sciences, Choua\"ib Doukkali University, PO Box 20, 24000 El Jadida, Morocco}
	\affiliation{Canadian Quantum  Research Center,
		204-3002 32 Ave Vernon,  BC V1T 2L7,  Canada}
	\author{El Houssine Atmani}
	\affiliation{Laboratory of Nanostructures and Advanced Materials, Mechanics and Thermofluids, Faculty of Sciences and Techniques, Hassan II University, Mohammedia, Morocco}
	
	
	\begin{abstract}
We investigate spin-dependent transmission in a curved graphene superlattice of $N$ cells where each one is made up of four regions. The first is concave, and the third is convex, two arcs of circles separated by a distance  $d$ from flat graphene sheets. The tunneling analysis allows us to determine all transmission and reflection channels associated with our system. As a result, we show that the number of cells acts by decreasing the transmissions with the same spin. We predict a solid spin-filtering effect when $d$ and $N$ are sufficiently large. Finally, it is determined that the degree and duration of suppression of the transmissions with the same spin over a range of energy are controllable using $d$.
	\end{abstract}

	\maketitle

	\section{Introduction}
	
Graphene is a carbon atom arranged in a single, thin, two-dimensional layer, like a honeycomb. Due to its unique band structure with band-tuning ability, graphene can therefore be likened to a gapless semiconductor. We thus speak of graphene as a semi-metallic material (in a semi-metal, the valence band and the conduction band overlap slightly, and the Fermi level falls in this zone, where the density of electronic states is low ). Graphene, being a zerogap semiconductor, has a band structure described by a linear dispersion relation at low energy, similar to massless Dirac-Weyl fermions \cite{saito1998,castro2009}. These properties are modular and can be combined endlessly. However, graphene properties are effective in the planar direction only, limiting the scope of graphene and applications. The properties of the carbon atom, in combination with the physical effects that occur in the layer, give rise to a number of unique properties. The curved $\pi$-conjugation of the carbon atoms in the lattice determines their unique chemical and physical behavior \cite{xlu}. Thanks to its exceptional thermal conductivity and functional reactivity, graphene can make important contributions to corrugated graphene (ripple) production.

Different techniques have been used to produce corrugated graphene, mainly electric arc discharge \cite{kane2020,khrabry2019}, and \cite{lijima}, which led to the first discovery of this class of materials. Indeed, it has become possible to create undulations by means of electrostatic manipulation \cite{alyobi2019}. Also, the ripples have been found to act as potential barriers for charged transporters that lead to their localization \cite{vasic2016}. Periodic ripples could be created using chemical vapor deposition \cite{gxni2012,zhang2014} or by thermal treatment \cite{bao,Miniussi}. The curvature of the graphene sheet surface affects the  $\pi$ orbitals which determine the electronic properties of graphene. This results in an improved spin-orbit coupling that could serve as a source of spin scattering \cite{ando,hurtas,delvalle}.
	In quantum mechanics, the study of electron movement in corrugated periodic chains provides an interesting tool to understand and get an idea of various physical properties of materials. There are several studies, such as the scattering of electrons through a periodically repeated chain \cite{andreas,puldak2020,smotlacha}, which demonstrate that electron transport through such a system is spin-dependent. Also, by considering a chain formed of concave and convex undulations to create a sheet of sinusoidal type graphene, the  electron scattering through the superlattice (chiral spin guide) was investigated. \cite{pudlak2015cooperative}. As a result, the spin filtering effect, which is very small in a single corrugation, becomes important with several corrugations. 
	We previously demonstrated that corrugated graphene subjected to an external mass term modifies the tunneling effect \cite{jaouad2021,jaouad2022}. In particular, it is found that resonances appear in reflection with the same spin, thus backscattering with a spin-up/-down is not null in ripple and transmission filtering becomes crucial. Also, spatial shifts for the total conduction are observed in our model, and the magnitudes of these shifts can be efficiently controlled by adjusting the band gap.

	We investigate electron tunneling through a corrugated graphene-based superlattice composed of $N$ elements, as shown in Fig. \ref{fig1}. A superlattice cell is made up of one concave ripple connected to a flat graphene piece with a distance  $d$ and another convex ripple connected to another flat graphene piece. The ripples in this case are identical and are made up of a curved graphene surface in the shape of a circle arc with radii $r_0$ and angles $\phi$. In the effective mass approximation \cite{ando,demartino}, we study the transmission of an electron flow through a rippled system by supposing that the incident electrons move along the $x$-axis (Fig. \ref{fig1}). We will consider a sufficiently wide graphene superlattice $W \gg L$, where $W$ and $L$ are the width and length of the graphene superlattice along the $y$-axis, respectively. We keep translational invariance along the $ y $-axis and neglect side effects.	
We derive the energy spectrum and use the transfer matrix to analytically calculate the different transmission channels. We show that the distance $d$ has an effect on the behavior of transmissions with up/down spin, and that the spin filtering effect is instantaneous, in contrast to the case where $d=0$, where the spin filtering effect survives a certain amount of energy. We notice that the transmission of electrons is extremely suppressed through the superlattice for certain energy values. Moreover, we prove that increasing $d$ causes a change in the duration of transmission suppression by decreasing the interval between peaks. Overall, we demonstrate that one of the most important tools to control the electronic transport properties of our system is to choose a graphene-based multiple superlattice, ripple (concave, convex), and flat graphene.

The paper is structured as follows. In Sec. \ref{sec2}, we set the mathematical tools and determine the solutions of the energy spectrum for a given cell.  The generalization to $N$-cell can easily be obtained since they are identical. Using the boundary conditions together with the matrix transfer approach,  we compute the transmissions of the states with different configurations of spin in Sec. \ref{sec3}. Sec. \ref{sec4} will be devoted to the numerical analysis and discussions of the obtained results to emphasize their basic features. We summarize the main conclusions in Sec. \ref{sec5}.

\section{Theoretical model}\label{sec2}

\begin{figure}[H]
	\begin{center}
		\includegraphics[scale=0.55]{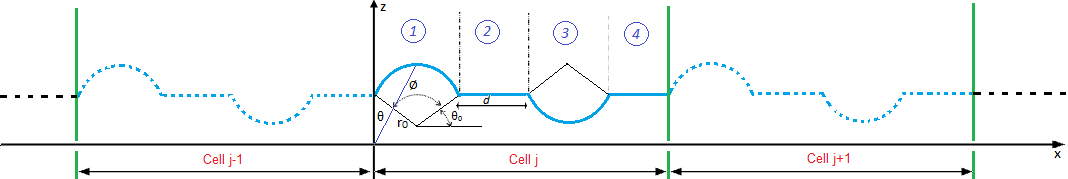}
	\end{center}
	\caption{Graphene superlattice composed of $N$ cells, each of which is made up of a juxtaposition of four regions.  }
\label{fig1}             
\end{figure}

As shown in Fig.~\ref{fig1}, we consider an infinitely large corrugated graphene. It is made up of $ N $ cells denoted by $j\ (j=0, \cdots, N-1) $, each of which is made up of four regions: a concave arc of a circle with radius $r_0$ and angle $\phi_j$, a flat graphene with length $d$, and another convex arc of a circle. The $j$-th elementary cell has three internal connection points: $x_2j=2r_0\sin\phi_j/2+j D$, $x_3j=x_2j+d$, $x_4j=4r_0\sin\phi_j/2+d$, and two extreme connection points: $x_1j=(j+1) D) $. One cell has a width of $D = 2d + 4r_0\sin\phi_j/2$. 

Based on pseudo-spin and spin, it is possible to write the Hamiltonian describing the aforementioned system as \cite{ando,puldak2014,jaouad2021}
	\begin{equation}\label{eq1}
		H=
		\begin{pmatrix}
			0 &  \gamma\left({p}_{x}-i {p}_{y}\right) \mathbb{I}+i \frac{\delta \gamma'}{4 r_0} {\sigma_{r}}-\frac{2 \delta \gamma p}{r_0} {\sigma_{y}}\\
			\gamma\left({p}_{x}+i {p}_{y}\right) \mathbb{I}-i \frac{\delta \gamma'}{4 r_0} {\sigma_{r}}-\frac{2 \delta \gamma p}{r_0} {\sigma_{y}} & 0
		\end{pmatrix}
	\end{equation}
where  ${p}_{x}=- \frac{i}\hbar {r_0}{\partial}_{\theta}$ and  ${p}_{y}=-i\hbar {\partial}_{y}$ are the momenta. The Pauli matrices  are $ {\sigma}_{x, y, z} $ with the rotating ones
\begin{align}
	{\sigma_{r}}={\sigma_{x}} \cos \theta-{\sigma_{z}} \sin \theta,
	\qquad 
{\sigma_{\theta}}=-{\sigma_{x}} \sin \theta-{\sigma_{z}} \cos \theta.
\end{align} 
The set of parameters    is  
  $ \gamma=-\frac{\sqrt{3}}{2} V_{p p}^{\pi} a =\gamma_{0} a, 	
	\gamma^{\prime}=\frac{\sqrt{3}}{2}\left(V_{p p}^{\sigma}-V_{p p}^{\pi}\right) a=\gamma_{1} a, p=\frac{1-3 \gamma^{\prime}} {8\gamma}  $ \cite{ando}.
	$V_{p p}^{\alpha}$ and $V_{p p}^{\pi}$, respectively, are the transfer integrals for the $\sigma$ and $\pi$ orbitals. The length of the primitive translation vector
	  in flat graphene is $a=$ $\sqrt{3} d \simeq 2.46\ \angstrom$ and $d$ is the distance between atoms in the unit cell. Spin-orbit coupling's intrinsic source, $\delta$, is provided by 
	\begin{align}
		\delta=i \frac{ \hbar}{4 m_{e}^{2} c^{2}\epsilon_{\pi \sigma}}\left\langle x\left|\frac{\partial V}{\partial x} {p}_{y}-\frac{\partial V}{\partial y} {p}_{x}\right| y\right\rangle	
	\end{align}
such that $V$ is  the atomic potential  and  $\epsilon_{\pi \sigma}=\epsilon_{2 p}^{\pi}-\epsilon_{2 p}^{\sigma}$, with  the energy $\epsilon_{2 p}^{\sigma}$   of $\sigma$ orbitals (localized between carbon atoms) and  the energy $\epsilon_{2 p}^{\pi}$   of $\pi$ orbitals (directed perpendicular to the curved surface). We will utilize the values $\gamma=6.39\ \angstrom\ \electronvolt, \gamma^{\prime}=17.04\ \angstrom\ \electronvolt, \delta=0.01$,  $p=0.1$ for our numerical calculations.	
In order to simply diagonalize $H$, we use a unitarty transformation in terms of $\sigma_y $ to remove the $\theta$-dependence. This is
\begin{equation}
	{U}=
	\begin{pmatrix}
		e^{i \frac{\theta}{2} {\sigma}_{y}} & 0 \\
		0 & e^{i \frac{\theta}{2} {\sigma}_{y}}
	\end{pmatrix}.
\end{equation}
It is clear that $U$ is a rotation matrix that rotates points in the $(x,z)$ plane in the opposite direction of the clock through the angle $\theta$. As a result, \eqref{eq1} becomes the Hamiltonian shown below
\begin{equation}
	\mathcal{H}={U} {H} {U}^{-1}.
\end{equation}

To deal with our task, we start by determining solutions of the energy spectrum. Indeed, in regions $1$ and $3$, the Hamiltonian takes the form 
\begin{equation}\label{eqq8}
	\mathcal{H}_{1,3}^{j}
	= 
	\begin{pmatrix}
		0 & 0 & -\gamma\partial_{y}-\mathrm{i} \frac{\gamma}{r_{0}} \partial_{\theta_{j}} & \mathrm{i}\left(\lambda_{y}+\lambda_{x}\right) \\
		0 & 0 & \mathrm{i}\left(\lambda_{y}-\lambda_{x}\right) &-\gamma\partial_{y}-\mathrm{i} \frac{\gamma}{r_{0}} \partial_{\theta_{j}} \\
		\gamma\partial_{y}-\mathrm{i} \frac{\gamma}{r_{0}} \partial_{\theta_{j}} & -\mathrm{i}\left(\lambda_{y}-\lambda_{x}\right) & 0 & 0 \\
		-\mathrm{i}\left(\lambda_{y}+\lambda_{x}\right) & \gamma\partial_{y}-\mathrm{i} \frac{\gamma}{r_{0}} \partial_{\theta_{j}} & 0 & 0
	\end{pmatrix}.
\end{equation}
We show that 
the four energy bands  are given by
\begin{equation}\label{eqq17}
		E_{js}^{s'}=s'\sqrt{t_{m_{j}}^{2}+t_{y}^{2}+\lambda_{x}^{2}+\lambda_{y}^{2}+2s \sqrt{t_{m_{j}}^{2}\left(\lambda_{x}^{2}\right)+\left(t_{y}^{2}+\lambda_{x}^{2}\right) \lambda_{y}^{2}}}
\end{equation}
and different quantities read as 
\begin{align}
	t_{m_{j}}=\frac{\gamma}{r_0} m_{j},\qquad  t_{y}=\gamma k_{y},\qquad  \lambda_{x}=\frac{\gamma}{2 r_0}(1+4 \delta p),\qquad   \lambda_{y}=\frac{\delta \gamma^{\prime}}{4 r_0}
\end{align}
where  $ s,s'=\pm $.
Then at normal incidence
($k_{y}=0$), \eqref{eqq17} reduces to
\begin{equation}\label{eqq18}
	E_{js}^{s'}=s'\sqrt{t_{m_{j}}^{2}+\lambda_{y}^{2}}+s\lambda_{x}
\end{equation}
showing the angular momentum 
\begin{equation}\label{eqq4}
	m_{js}^{s'}=s'\frac{r_{0}}{\gamma} \sqrt{\left(E_{js}^{s'}-s \lambda_{x} \right)^{2}-\lambda_{y}^{2}} 
\end{equation}
and we set $m_{js}^{+}=m_{js}$ after observing that 
$m_{js}^{+}=-m_{js}^{-}$. 
As solutions corresponding to four bands \eqref{eqq18}, we obtain 
the eigenspinors 
\begin{equation}\label{sp13}
	\Psi_{1,3}^{j,s,s'}(\theta_{j}, m_{js})=\left(\mathbb{I} \otimes e^{-i \frac{\sigma_{y}}{2} \theta_{j}}\right)
	\begin{pmatrix}
		s i  t_{m_{js}} \\
		-s\lambda_{y}+\left(E_{js}^{s'}-s\lambda_{x}\right)\\
		- i (\lambda_{y}-(s E_{js}^{s'}-\lambda_{x}))\\
		t_{m_{js}}
	\end{pmatrix}
	e^{i m_{js} \theta_{j}}
\end{equation}
where $ e^{i m_{js} \theta_{j}} $ are eigenstates of the angular momentum $-i\partial_{\theta_{j}}$ associated with the integers eigenvalues $m_{js}$.
Finally, in regions $1 $ and $3$ the total eigenspinors is given by
\begin{equation}
	\Psi_{1,3}^{j}(\theta_{j})=	a_{+j} \Psi_{1,3}^{j,+,+}\left(\theta_{j}, m_{j+}\right)
	+b_{+j} \Psi_{1,3}^{j.+,-}\left(\theta_{j}, -m_{j+}\right)+a_{-j} \Psi_{1,3}^{j,-,+}\left(\theta_{j}, m_{j-}\right)
	+b_{-j} \Psi_{1,3}^{j,-,-}\left(\theta_{j},-m_{j-}\right)
\end{equation}
with	$a{_{\pm}}_{j}$ and $ b{_{\pm}}_{j},$ denote the coefficients of the linear combination corresponding to each component derived from \eqref{sp13}. It is convenient for our task to write $\Psi_{1,3}^{j}(\theta_{j})$ in  matrix form
\begin{equation}\label{sp132}
	\Psi_{1,3}^{j}(\theta_{j})=	M_{1,3}^{}(\theta_{j})
	\left(e^{\mathrm{i} {m_{js}}\theta_{j}}\right) 
	D_{1,3j}
\end{equation}
and all matrices are established as
\begin{align}
	&	M_{1,3}({\theta_{j}})=\Lambda(-\theta_{j}) M_{1,3}
	\\
	&	\Lambda(\theta_{j})=\begin{pmatrix}
		\cos \frac{\theta_{j} }{2} & -\sin  \frac{\theta_{j}  }{2} & 0 & 0 \\
		\sin \frac{\theta_{j}  }{2} & \cos  \frac{\theta _{j} }{2} & 0 & 0 \\
		0 & 0 & \cos \frac{\theta _{j} }{2} & -\sin  \frac{\theta _{j} }{2} \\
		0 & 0 & \sin  \frac{\theta _{j} }{2} & \cos  \frac{\theta _{j} }{2}
	\end{pmatrix}
	\\
	& \label{eq26}
	{M}_{1,3} =\begin{pmatrix}
		-i t_{m_j+} & -i t_{m_j-} & i t_{m_j+} & i t_{m_j-} \\
		\lambda_y-\sqrt{\lambda_y^2+t_{m_j+}^2} & \lambda_y+\sqrt{\lambda_y^2+t_{m_j-}^2} & \lambda_y-\sqrt{\lambda_y^2+t_{m_j+}^2} & \lambda_y+\sqrt{\lambda_y^2+t_{m_j-}^2} \\
		i \left(\lambda_y-\sqrt{\lambda_y^2+t_{m_j+}^2}\right) & -i \left(\lambda_y+\sqrt{\lambda_y^2+t_{m_j-}^2}\right) & i
		\left(\lambda_y-\sqrt{\lambda_y^2+t_{m_j+}^2}\right) & -i \left(\lambda_y+\sqrt{\lambda_y^2+t_{m_j-}^2}\right) \\
		-t_{m_j+} & t_{m_j-} & t_{m_j+} & -t_{m_j-} \\
	\end{pmatrix}\\
	&
	\left(e^{\mathrm{i} {m_{js}}\theta_{j}}\right) 
	=	\text{diag}\left(e^{\mathrm{i} m_{j+} \theta_{j}}, e^{\mathrm{i} m_{j-} \theta_{j}}, e^{\mathrm{-i} m_{j+} \theta_{j}}, e^{\mathrm{-i} m_{j-} \theta_{j}}
	\right)
	\\
	&
	D_{1,3j}=\begin{pmatrix}
		a_{+j} \\
		b_{+j}\\
		a_{-j} \\
		b_{-j}
	\end{pmatrix}.
\end{align}

Regarding   the flat graphene regions $2 $ and $4$, the solutions of the energy spectrum can be easily obtained. Indeed, at $k_y=0$, we derive the following two bands 
\begin{equation}\label{fen}
		E_{j\tau}=\tau \gamma|k_{j}|
\end{equation}
associated to the eigenspinors 
\begin{equation}\label{eqq20}
	\Psi_{2,4}^{j,\tau,\kappa}(x, k_{j})=\frac{1}{2}
	\begin{pmatrix}
		\tau \\
		1
	\end{pmatrix} \otimes
	\begin{pmatrix}
		1 \\
		\kappa i
	\end{pmatrix}
	e^{i k_{j}  x}
\end{equation}
with $ \kappa=\pm$, $ k_{j}=k_{x} $ and    $\tau=\text{sgn}(k_{j})=\pm$ refers to  conductance and valence bands, respectively. The total eigenspinors can therefore be written as 
\begin{equation}
	\psi_{2,4}^{j}(x, k_{j})=\alpha_{j}\psi_{2,4}^{j,+,+}(x, k_{j})+\beta_{j}\psi_{2,4}^{j,-,+}(x, k_{j})+\gamma_{j} \psi_{2,4}^{j,+,-}(x,-k_{j})+\xi_{j} \psi_{2,4}^{j,-,-}(x,-k_{j})
\end{equation}
where $\alpha_{j}, \beta_{j}$, $\gamma_{j}$ and $\xi_{j}$ indicate the coefficients of the linear combination associated with each incident wave \eqref{eqq20} and its reflected one. It can be expressed in the following matrix form
\begin{equation}\label{sp242}
	\psi_{2,4}^{j}(x,k_{j})=	M_{2,4}  \left(e^{\mathrm{i} {K_{j}} x}\right)
	D_{2,4j}
\end{equation}
where different matrices are set to be
\begin{align}
	&M_{2,4} =\begin{pmatrix}
		1 & 1 & -1 & -1 \\
		i  & -i & -i & i \\
		1 & 1 & 1 & 1 \\
		i & -i & i & -i
	\end{pmatrix}\\
& \left(e^{\mathrm{i} {K_{j}} x}\right)=\text{diag}\left(
e^{i k_{j}  x }, e^{i k_{j}  x }, e^{-i k_{j}  x }, e^{-i k_{j}  x }
\right) 
	\\
	&
	D_{2,4j}=\begin{pmatrix} 
		\alpha_{j} \\
		\beta_{j}\\
		\gamma_{j} \\
		\xi_{j}
	\end{pmatrix}.
\end{align}

In the next, we will see how to use the above results to study the tunneling effect. More precisely, we will be interested in  different transmission channels and their behaviors under various conditions of the physical parameters. Also, we will discuss and emphasize what is new with respect to literature, and in particular, the interesting results obtained in \cite{Busa2019}.

	\section{Transmission channels}\label{sec3}

We begin by calculating the transmission channels of a single cell $j$ using the continuity of eigenspinors \eqref{sp132} and \eqref{sp242} in conjunction with the transfer matrix method.  
For simplicity, we remove the subscript $j$, and then at the contact points of different regions, we write
\begin{align}
&	M_{2}D_{in}=M_{1}\left(-\frac{\phi}{2}\right) \left(e^{-\mathrm{i} {m} \frac{\phi}{2}}\right)
	D_{1 } 
\\
&
	M_{1}\left(\frac{\phi}{2}\right)
\left(e^{\mathrm{i} {m} \frac{\phi}{2}}\right) 
D_{1}=M_{2} \ 
\left(e^{\mathrm{i} {K} x_{2} }\right) 
D_{2} 
\\
&
	M_{2}\ 
\left(	e^{\mathrm{i} {K} x_{3}}\right) D_{2}=M_{1}\left(\frac{\phi}{2}-\pi\right) 
\left(	e^{\mathrm{i} {m} (\frac{\phi}{2}-\pi)}\right) D_{3}
\\
&
	M_{1}\left(\pi-\frac{\phi}{2}\right)
\left(e^{\mathrm{i} {m}( \pi-\frac{\phi}{2})}\right) 
D_{3}=M_{2}D_{out}
\end{align}
where the input $D_{{in}}$ and the output $D_{{out}}$ of 
are given by
\begin{equation}\label{eq22}
	D_{in}=\begin{pmatrix}
		\alpha \\
		\beta\\
		r_{\uparrow}^{\xi} \\
		r_{\downarrow}^{\xi}
	\end{pmatrix},\qquad
	D_{out}=\begin{pmatrix}
		t_{\uparrow}^{\xi} \\
		t_{\downarrow}^{\xi} \\
		0 \\
		0
	\end{pmatrix}.
\end{equation}
As well as ($ \alpha=0, \beta=1$) for the spin down polarization with (${\xi}={\downarrow}$) and ($\alpha=1, \beta=0$) for the spin up polarization with ($ \xi=\uparrow$) following the electron flux displacement.
According to the previous arguments, we construct the transfer matrix between $D_{in}$ and $D_{out}$:
\begin{align}
		D_{in}&=- M_{2}^{-1}M_{1}\left(-\frac{\phi}{2}\right)	\left(e^{\mathrm{-i} {m}\phi}\right)  M_{1}^{-1}\left(\frac{\phi}{2}\right) M_{2}  \left(e^{\mathrm{-i} {K} d}\right)M_{2}^{-1} M_{1}\left(\frac{\phi}{2}-\pi\right)\left(e^{\mathrm{i} {m}\phi}\right)  M_{1}^{-1}\left(\pi-\frac{\phi}{2}\right) M_{2} 
		D_{out}\notag\\
		&=\Omega  D_{out}	
\end{align}
where the matrix $\Omega$ is
\begin{equation}\label{eqq19}
	\Omega=	\left(\begin{array}{cccc}
		z_{1} & 0 & 0 & z_{1}^{\prime} \\
		0 & z_{2} & z_{2}^{\prime} & 0 \\
		0 & z_{2}^{\prime *} & z_{2}^{*} & 0 \\
		z_{1}^{\prime *} & 0 & 0 & z_{1}^{*}
	\end{array}\right)
\end{equation}
and we have set
\begin{align}
&	z_{1} =	\frac{2 {\lambda_y}^2 \cos (d k) \sin ^2\left(m_- \phi \right)}{t_{m_-}^2}+e^{-i d k}
\\
&
	z_{2}=  \frac{2 {\lambda_y}^2 \cos (d k) \sin ^2\left(m_+ \phi \right)}{t_{m_+}^2}+e^{-i d k}
\\
&
	z_{1}^{\prime} =	\frac{2 {\lambda_y} e^{\frac{i \phi }{2}} \cos (d
		k) \sin \left(m_- \phi \right) \left(\sqrt{{\lambda_y}^2+t_{m_-}^2} \sin \left(m_- \phi \right)+i t_{m_-} \cos \left(m_- \phi
		\right)\right)}{t_{m_-}^2} 
\\
&
	z_{2}^{\prime}= - \frac{2 {\lambda_y} e^{-\frac{i \phi }{2}} \cos (d k)
		\sin \left(m_+ \phi \right) \left(\sqrt{{\lambda_y}^2+t_{m_+}^2} \sin \left(m_+ \phi \right)+i t_{m_+} \cos \left(m_+ \phi \right)\right)}{t_{m_+}^2}.
\end{align}

At this level, we can use the above results to explicitly determine  different transmission channels. Indeed, we show them to be  
\begin{align}
&	\label{eq30}
		T_{\uparrow}^{\uparrow}=\left|t_{\uparrow}^{\uparrow}\right|^{2}=\frac{1}{\left(1+2\left(z_{-}\right)^{2}\right)^2 \cos^2(d k)+\sin^2(dk)}
\\
&\label{eq31}
		T_{\downarrow}^{\downarrow}=\left|t_{\downarrow}^{\downarrow}\right|^{2}=\frac{1}{\left(1+2\left(z_{+}\right)^{2}\right)^2 \cos^2(d k)+\sin^2(dk)}
\end{align}
such that $ z_{s} $ is 
\begin{equation}
	z_{s}=\frac{\lambda_{y} \sin \left(m_{s} \phi\right)}{t_{m_{s}}},
	\qquad s=\pm. 
\end{equation}
Let us test the validity of our results by taking a limiting case. Indeed, for $d = 0$ (without flat graphene), the transmission obtained in \cite{pudlak2015cooperative} is recovered
\begin{align}	T_{\uparrow}^{\uparrow}=\frac{1}{\left(1+2(z_{-})^{2}\right)^{2}}, \qquad
		T_{\downarrow}^{\downarrow}=\frac{1}{\left(1+2(z_{+})^{2}\right)^{2}}.
\end{align}

 To generalize transmissions to $ N $ cells, we use the set of equations  (\ref{eq22}-\ref{eqq19}). Indeed, when we combine $\Omega$ matrices for $N$ connected cells 
 corresponding to spin-up, we get 
\begin{align}\label{eq35}
	\begin{pmatrix}
		1 \\
		r_{\downarrow}^{\uparrow}
	\end{pmatrix}&=\begin{pmatrix}
		A_{11} & A_{12} \\
		A_{21} & A_{22}
	\end{pmatrix}^{N} \begin{pmatrix}
		t_{\uparrow}^{\uparrow}	 \\
	0
	\end{pmatrix}\\
& \label{NN} =\begin{pmatrix}
	N_{11} & N_{12} \\
	N_{21} & N_{22}
\end{pmatrix}\begin{pmatrix}
t_{\uparrow}^{\uparrow}	 \\
0
\end{pmatrix}
\end{align}
and the matrix elements are given by
\begin{align}\label{eq36}
	A_{11}=	z_{1}=	A_{22}^{*}, \qquad 
	A_{12}=	z_{1}^{\prime}=A_{21}^{*}.
\end{align}
Knowing that there exists the unitary transformation $U$, then we can  diagonalize the last matrix in \eqref{NN}. As a result, we have the following
\begin{equation}\label{eq38}
	U^{-1} \begin{pmatrix}
		N_{11} & N_{12} \\
		N_{21} & N_{22}
	\end{pmatrix} U=\left(\begin{array}{cc}
		\lambda_{+}^- & 0 \\
		0 & \lambda_{-}^-
	\end{array}\right)
\end{equation}
giving rise to the elements
\begin{align}
	&\label{eq40}
	N_{11}=N_{22}^{*}=\frac{A_{11}\left(\left(\lambda_{+}^{-}\right)^{N}-
		\left(\lambda_{-}^{-}\right)^{N}\right)+
		\left(\lambda_{-}^{-}\right)^{N-1}-
		\left(\lambda_{+}^{-}\right)^{N-1}}{\lambda_{+}^{-}-\lambda_{-}^{-}}
\\
&\label{eq41}
	N_{12}=N_{21}^{*}=A_{12} \frac{\left(\lambda_{+}^{-}\right)^{N}-
		\left(\lambda_{-}^{-}\right)^{N}}{\lambda_{+}^{-}-\lambda_{-}^{-}}. 
\end{align}
 We demonstrate that the matrix's eigenvalues have the form 
\begin{equation}
	\lambda_{\pm}^{-}= \left(\frac{2 \lambda_y^2 \sin ^2\left(m_- \phi \right)}{t_{m_-}^2}+1\right)\cos (d k)\pm
	\sqrt{ \left(\frac{2 \lambda_y^2 \sin ^2\left(m_- \phi \right)}{t_{m_-}^2}+1\right)^2\cos^2 (d k) -1}
\end{equation}
and the condition $\left|N_{11}\right|^{2}-\left|N_{12}\right|^{2}=1$ is satisfied.  Therefore, we end up with the transmission with spin up for $N$ cells as follows
\begin{equation}
	T^{\uparrow}_{\uparrow}=\frac{1}{1+\left(2 z_{-}A_{N}^{-}\cos(k d) \right)^{2}(z_{-}^2+1)}
\end{equation}
where we have set $ A_{N}^{-}=\frac{\left(\lambda_{+}^{-}\right)^{N}-
	\left(\lambda_{-}^{-}\right)^{N}}{\lambda_{+}^{-}-\lambda_{-}^{-}} $, which corresponds to $s=-$.
To determine the transmission with spin-down for electron scattering with spin-down ($s=+$), we apply the same logic that was used with (\ref{eq35}-\ref{eq41}). This process yields to 
\begin{equation}
	T^{\downarrow}_{\downarrow}=\frac{1}{1+\left(2 z_{+}A_{N}^{+}\cos(k d) \right)^{2}(z_{+}^2+1)}
\end{equation}
where the eigenvalues are 
\begin{equation}
	\lambda_{\pm}^{+}= \left(\frac{2 \lambda_y^2 \sin ^2\left(m_+ \phi \right)}{t_{m_+}^2}+1\right)\cos (d k)\pm
	\sqrt{ \left(\frac{2 \lambda_y^2 \sin ^2\left(m_+ \phi \right)}{t_{m_+}^2}+1\right)^2\cos^2 (d k) -1}
\end{equation}
and $ A_{N}^{+} $ is defined in terms of $ \lambda_{\pm}^{+} $ as it has been done for $ A_{N}^{-} $.
At this level, there are some  remarks to state. Because \eqref{eqq4} shows $ m_+\neq  m_-$, we have $\lambda_{\pm}^{+}\neq \lambda_{\pm}^{-}$. It is worth noting that the presence of $\cos (d k) $ improves our transmissions by generalizing them and providing another controllable parameter to tune the tunneling effect. To recover the results obtained in \cite{pudlak2015cooperative}, we require $d= 0$ (without flat graphene) to end up with
\begin{equation}
	T_{\uparrow}^{\uparrow}=\frac{4}{\left(\left(C_{-}^{+}\right)^N+
	\left(C_{-}^{-}\right)^N\right)^2}, \qquad 
	T_{\downarrow}^{\downarrow}=\frac{4}
	{\left(\left(C_{+}^{+}\right)^N+\left(C_{+}^{-}\right)^N\right)^2}
\end{equation}
and the parameters $C_{s}^{\pm}$ have been defined as
\begin{equation}
	C_{s}^{\pm}=\sqrt{1+\left(z_{s}\right)^{2}} \pm z_{s}, \qquad s=\pm.
\end{equation}

\section{Results and Discussion}\label{sec4}

\begin{figure}[h]
	\centerline{
		\subfloat[]{
			\includegraphics[scale=0.5]{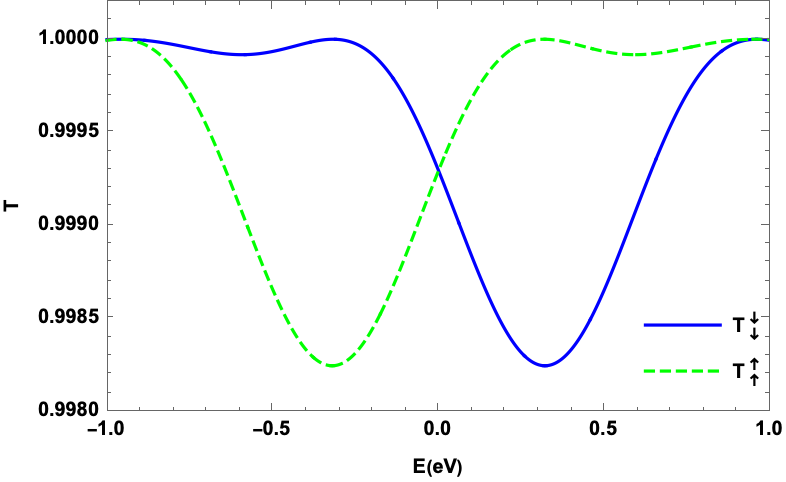}
			\label{subfigurea}}\hspace{2cm}
		{\subfloat[]{
				\includegraphics[scale=0.5]{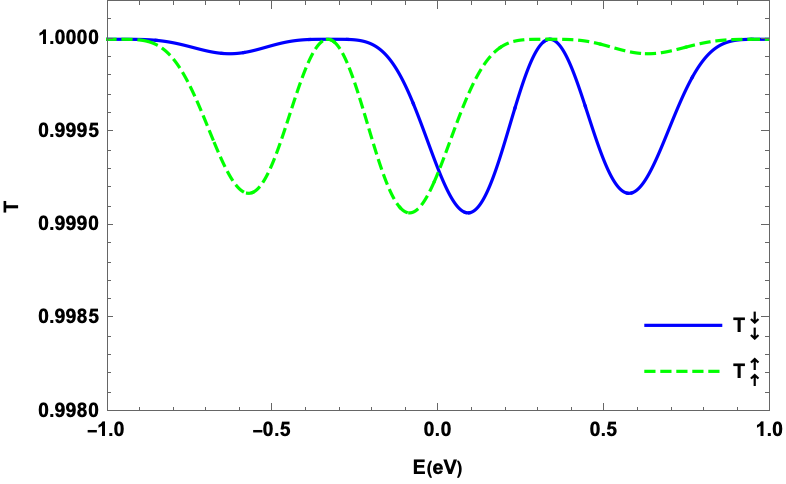}
				\label{subfigureb}}}}
	\centerline{
		\subfloat[]{
			\includegraphics[scale=0.5]{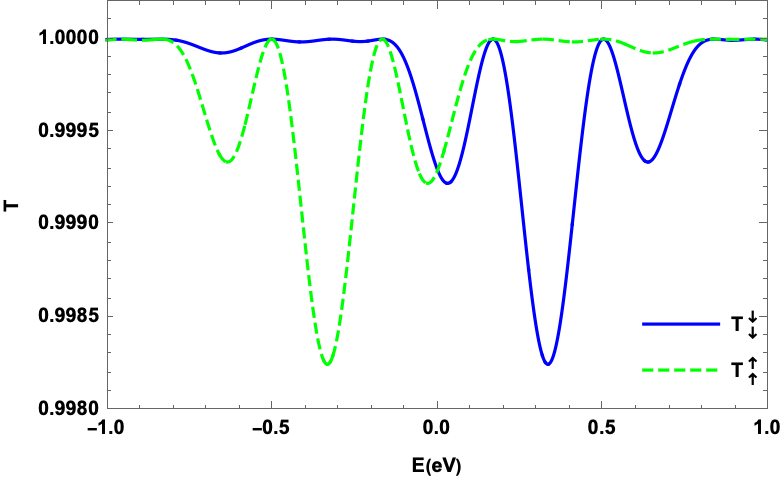}
			\label{subfigurec}}\hspace{2cm}
		{\subfloat[]{
				\includegraphics[scale=0.5]{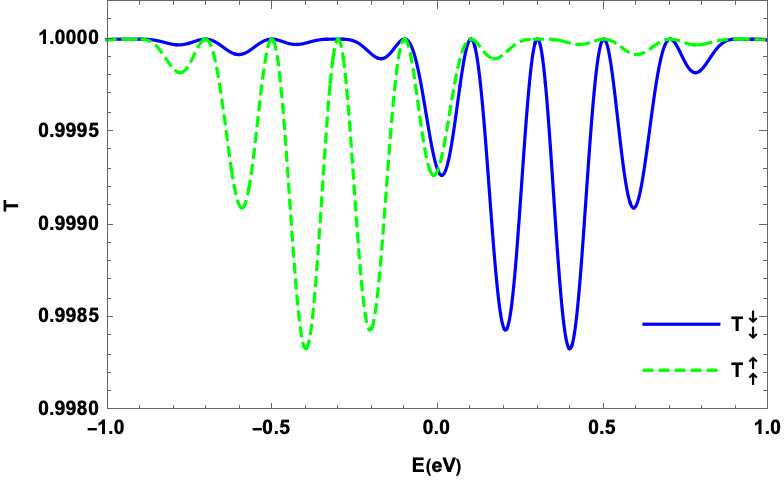}
				\label{subfigured}}}}
	\caption{(color online) Transmission channels $T_{\uparrow}^{\uparrow}$ (green) and $T_{\downarrow}^{\downarrow}$ (blue) 
		versus incident energy $E$ for $\phi=\pi$,  $r_{0}=10\ \text{\AA{}}$, $N = 1$ and
		four values of $d$ length of flat graphene     (a): $d=0$,  (b): $d=30\ \text{\AA{}}$, (c): $d=60\ \text{\AA{}}$, (d): $d=100\ \text{\AA{}}$. 
	}
	\label{fig2}
\end{figure}

We begin our discussions by establishing some interesting results that can be derived as limiting cases. Indeed, regardless of the physical parameters, both transmissions confirm the symmetry $T_{\uparrow}^{\uparrow}(-E)=T_{\downarrow}^{\downarrow}(E)$. The maximum transmission  $T_{\uparrow}^{\uparrow} = 1$ occurs under the conditions
\begin{align}
	kd=\left(l+\frac{1}{2}\right)\pi, \qquad m_{-} \phi=\pi n,\qquad l,n\in\mathbb{Z}
\end{align}
where $k$ is wave vector along $x$-axis, $d$ is length of flat graphene, $m_-$ is given in (\ref{eqq4}) and $\phi$ is an angle of  the circle arc. The same analysis can be applied to the transmission $T_{\downarrow}^{\downarrow}$ when $m_-$ is replaced by $m_+$. 
It is worth noting that, in addition to the condition obtained in \cite{smotlacha}, there is another constraint on the distance $d$, then we write 
\begin{equation}\label{dcon}
d=\frac{\pi}{k}\left(l+\frac{1}{2}\right), \qquad	\phi=\frac{\pi n \gamma}{r_{0} \sqrt{ (E_- +\lambda_{x})^{2}-\lambda_{y}^{2}}}.
\end{equation}
As we will see next, these quantities will play an important role in characterizing the transmission process for our system and therefore allow us to derive interesting results.

  We choose angle $\phi \approx \pi$ and radius  $r_{0}=10\ \text{\AA{}}$ of the circle arc for numerical purposes in addition to the parameters fixed in section \ref{sec2}. 
  In Fig. \ref{fig2}, we plot the two transmission channels as a function of incident energy $E$.
 It follows that, for $E > 0$, there is a high probability of transmission  for spin-up electrons from the left side of our superlattice, one obtains the same amplitude for the transmission probability for spin-down electrons from the right side, as shown in Fig.~\ref{subfigurea}.
 With $d\neq0$, the transmission changes behavior by creating oscillations and reaching minimums at various $E$ values in Fig.~\ref{subfigureb}. The interval between these minima becomes narrow.
In  Figs.~(\ref{subfigurec}, \ref{subfigured}), the oscillations multiply as $d$ increases, the interval between oscillations becomes smaller, but there is no great change in the amplitude of the transmission. As a result, the introduction of the distance d produces more oscillations than that without distance, but with a small variation between the spin-up and spin-down transmission amplitude for a single cell.

	\begin{figure}[H]
	\centerline{
		\subfloat[]{
			\includegraphics[scale=0.5]{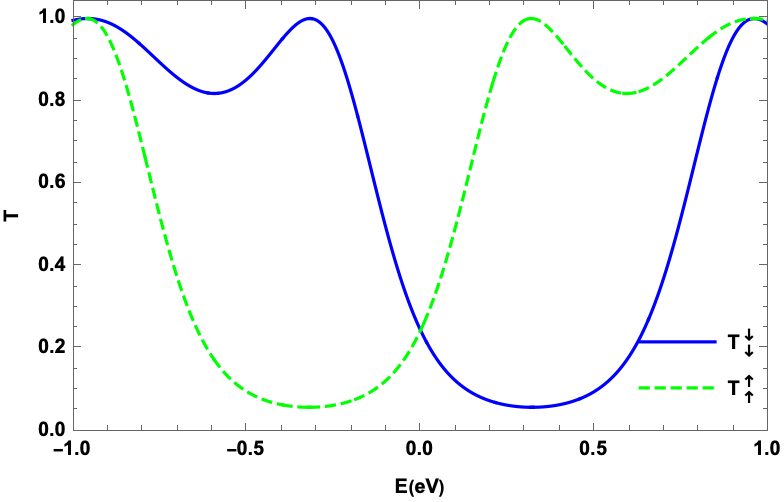}
			\label{subfigure3a}}\hspace{2cm}
		{\subfloat[]{
				\includegraphics[scale=0.5]{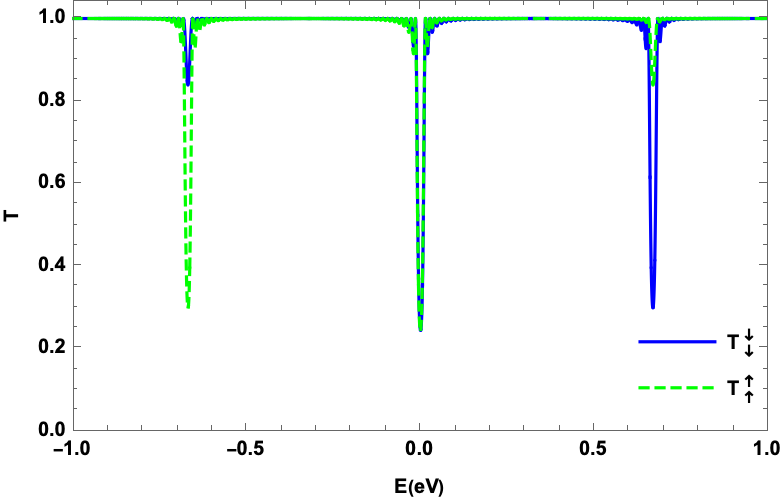}
				\label{subfigure3b}}}}
	\centerline{
		\subfloat[]{
			\includegraphics[scale=0.5]{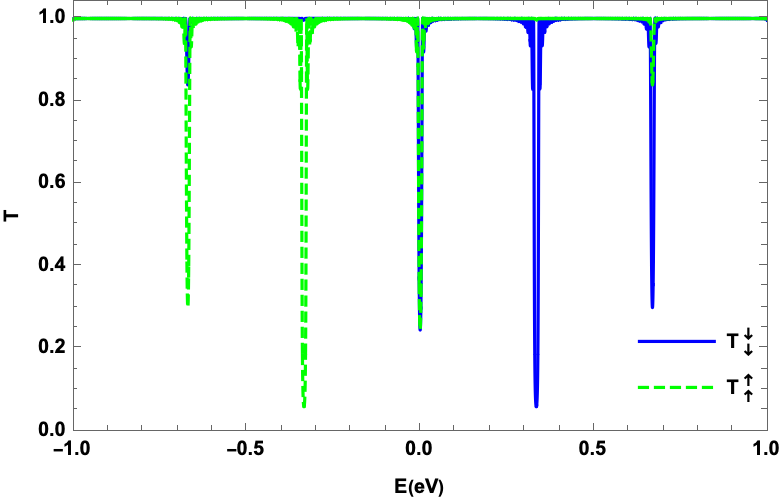}\hspace{2cm}
			\label{subfigure3c}}
		{\subfloat[]{
				\includegraphics[scale=0.5]{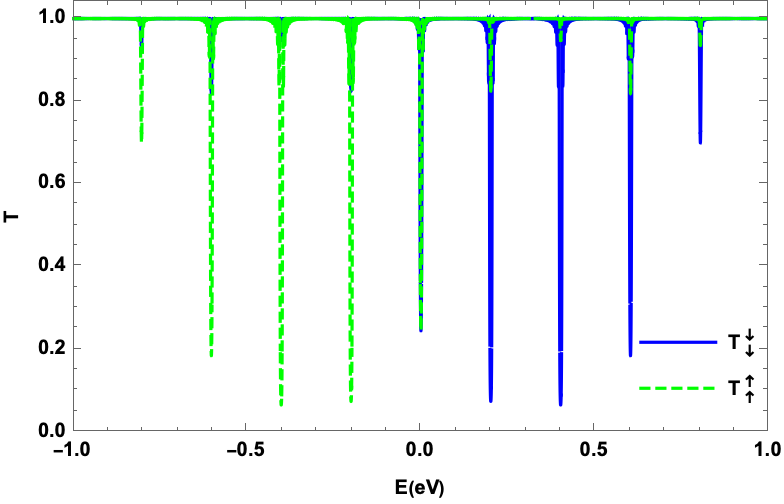}
				\label{subfigure3d}}}}
	\caption{(color online) 	The same as in Fig. \ref{fig2}, with
		$N = 50$ cells.}
	\label{fig3}
\end{figure}
In Fig.~\ref{fig3}, the transmission is presented in terms of energy for the same parameters of Fig.~\ref{fig2} but with a number of cells $N=50$.  
As shown in Fig.~\ref{subfigure3a},
at $d=0$, the transmission with spin down decreases continuously and approaches zero for an energy interval $E > 0$, whereas the transmission with spin up also decreases for $E< 0$. Transmission suppression persists for a certain energy interval. 
In Fig.~\ref{subfigure3b}, the behavior of the transmission is completely changed with the appearance of the peaks, and the transmission takes the minimum value. The spacing between the peaks is a little large. Increasing $d$ results in a greater number of peaks and a smaller spacing between them in Figs.~(\ref{subfigure3c}, \ref{subfigure3d}). From $N=50$, we notice that the suppression starts to occur with a smaller energy step, see (\ref{eq54}).

\begin{figure}[h]
	\centerline{
		\subfloat[]{
			\includegraphics[scale=0.5]{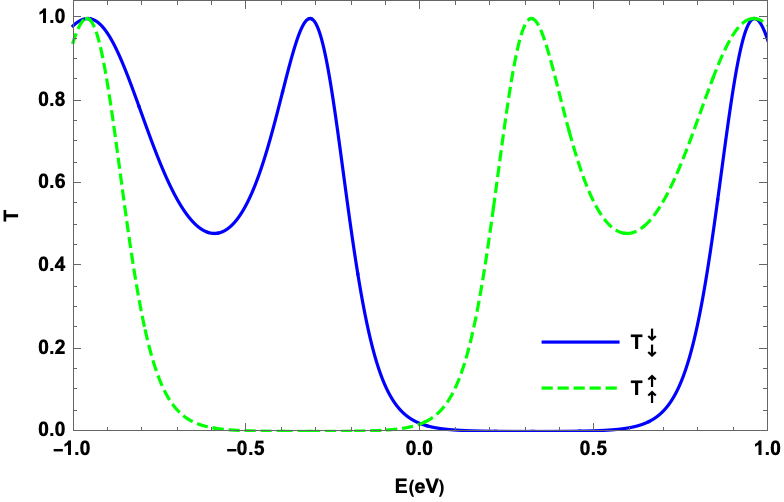}
			\label{subfigure4a}}\hspace{2cm}
		{\subfloat[]{
				\includegraphics[scale=0.5]{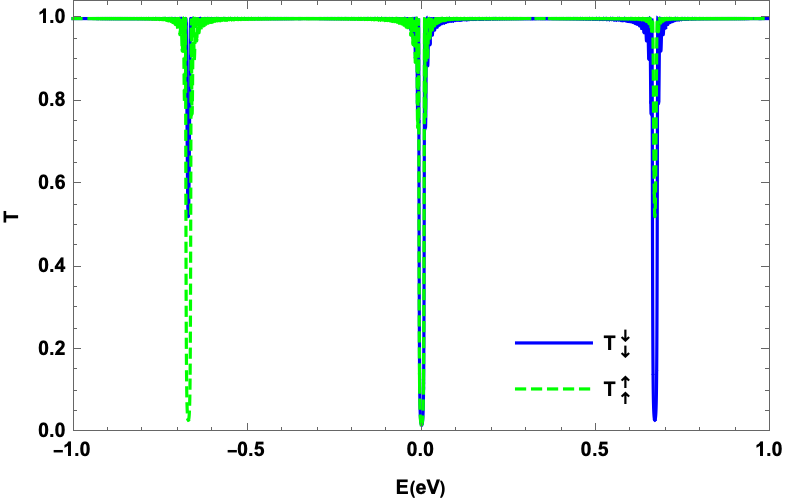}
				\label{subfigure4b}}}}
	\centerline{
		\subfloat[]{
			\includegraphics[scale=0.5]{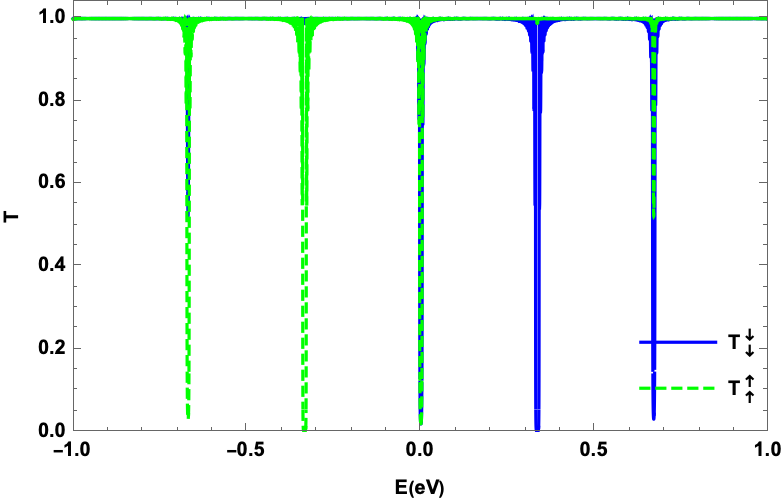}\hspace{2cm}
			\label{subfigure4c}}
		{\subfloat[]{
				\includegraphics[scale=0.5]{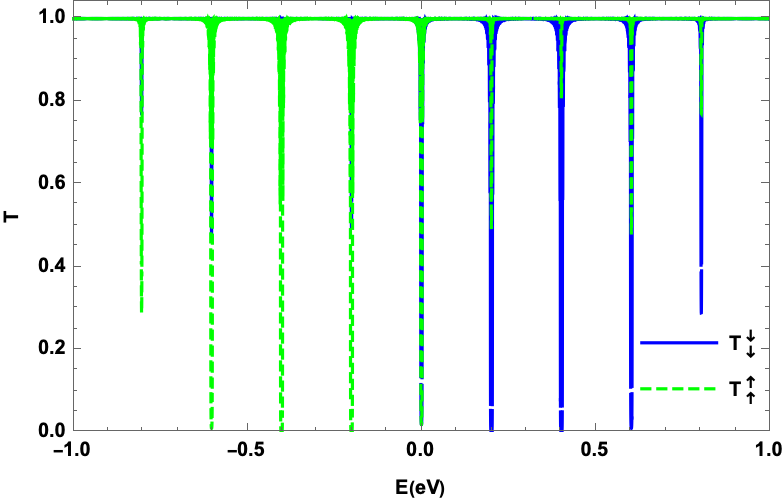}
				\label{subfigure4d}}}}
	\caption{(color online) 	The same as in Fig. \ref{fig2}, with
		$N = 100$ cells.}
	\label{fig4}
\end{figure}

Transmission is shown in terms of $E$ in
 Fig.~\ref{fig4}  for the same physical parameters as in Fig.~\ref{fig2}, but with $N=100$ cells. In  Fig.~\ref{subfigure4a}, we show that at $d=0$, the transmission with spin down decreases continuously and reaches zero value for a positive energy interval, likewise for the transmission with spin up for negative energy values. Transmission suppression survives for one energy interval. 
Increasing $d$ to $30\ \text{\AA{}}$ in Fig.~\ref{subfigure4b}, we observe that the suppression takes place in a discrete fashion and cancels out for each peak. 
  As shown in Figs.~(\ref{subfigure4c}, \ref{subfigure4d}), increasing the cell number $ N $ and distance $ d $ causes total suppression of transmission with a smaller energy step. The suppression effect becomes stronger as $N$ grows larger (with fixed $r_0$, $\phi$, and $d$).  As a result,  $d$  allows us to control the degree of suppression of  transmission with spin up/down in a range of energy, see (\ref{eq54}). Otherwise, increasing  $d$  gives rise to the periodic suppression of transmission channels with a small energy step, which can be explained using the strategy adopted in \cite{Busa2019}. 
  This outcome can be regarded as the quantum interference of the waves reflected by the external extremities of regions 1 and 2.
In fact, the wave reflected from the first interferes with the wave reflected from the second. The path length between the first and second interfering waves is $2(L_r+d)$, where $L_r$ is the ripple length. 
 If the following condition is met, the interference between the electrons reflected on the first and second boundaries is maximal: 

\begin{equation}
	k  (L_{r}+d)=\left( l+\frac{1}{2}\right)\pi, \qquad l\in \mathbb{Z}
\end{equation}
where $L_{r}\approx r_{0} \phi$ is the arc length. Because we have $E=\gamma k$, see \eqref{fen}, the energy can be quantized when transmission is suppressed, such that
\begin{equation}
	E_{l} = \frac{\gamma \pi}{r_{0} \phi+d}\left(l+\frac{1}{2}\right).
\end{equation}
This is equivalent to the energy of a harmonic oscillator, indicating that the system is oscillating at quantum interference with frequency $\frac{\gamma \pi}{r_{0} \phi+d}$. Therefore, we determine the energy step where the suppression occurs
as a gap
\begin{equation}\label{eq54}
	\Delta E=E_{l+1}-E_{l}=\frac{\gamma \pi}{r_{0} \phi+d}.
\end{equation}
These results are compatible with those shown in \eqref{dcon}
and then the maximum transmission occurs with the distance
\begin{equation}\label{eq56}
	d_{l}=\frac{\gamma \pi}{E}\left(l+\frac{1}{2}\right).
\end{equation}

\begin{figure}[H]
	\centerline{
		\subfloat[]{
			\includegraphics[scale=0.42]{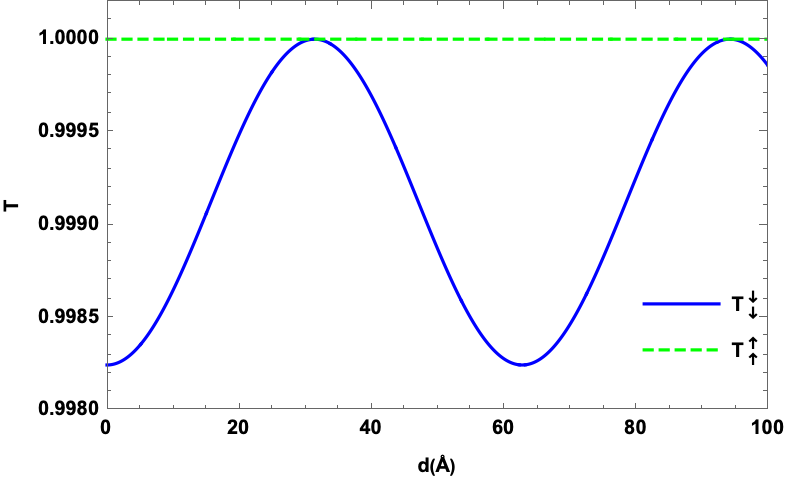}
			\label{subfigure5a}}
		{\subfloat[]{
				\includegraphics[scale=0.42]{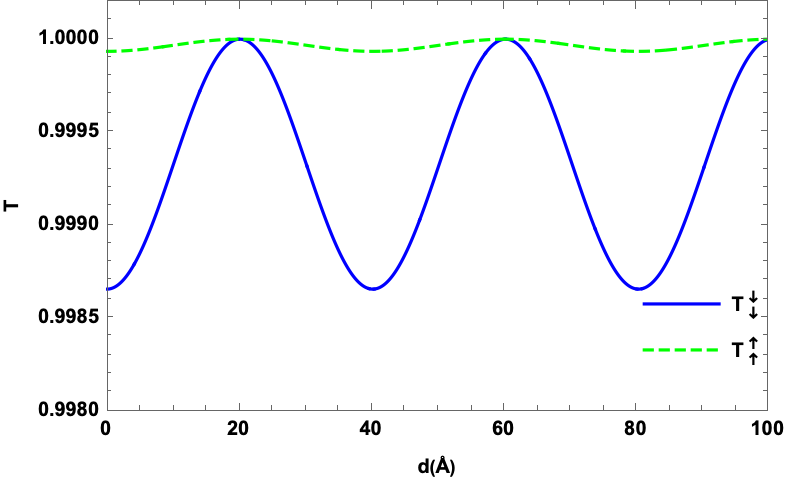}
				\label{subfigure5b}}
			{	\subfloat[]{
					\includegraphics[scale=0.42]{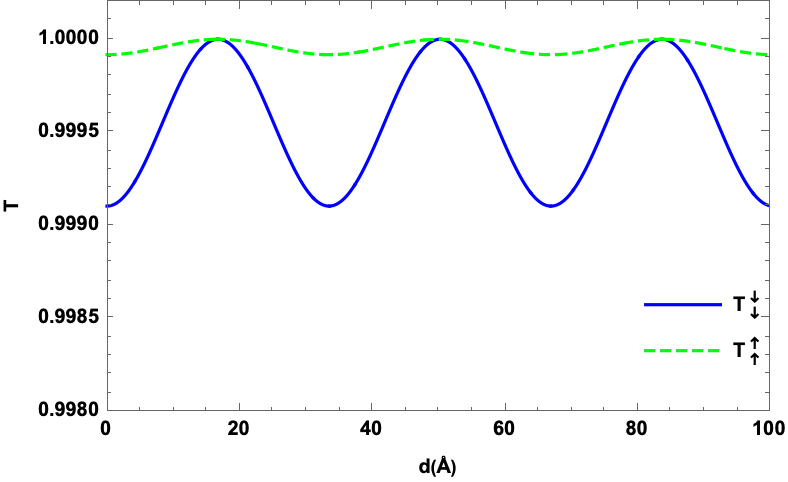}
					\label{subfigure5c}}}}}
	\centerline{
		\subfloat[]{
			\includegraphics[scale=0.42]{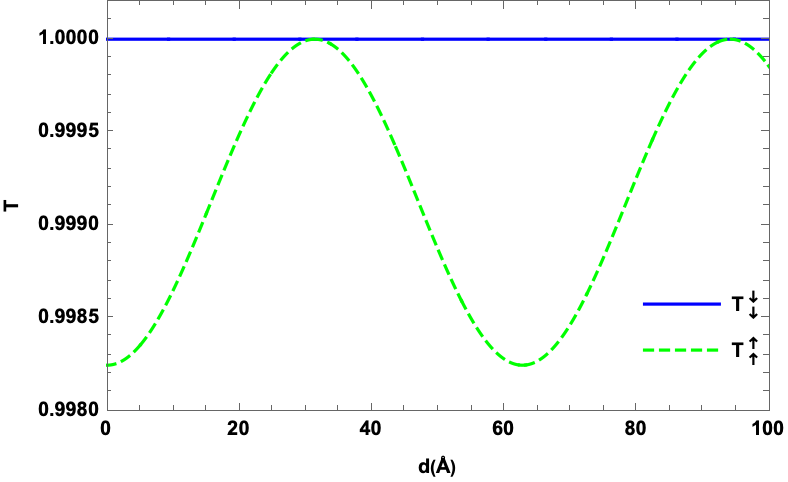}
			\label{subfigure5d}}{
			\subfloat[]{
				\includegraphics[scale=0.42]{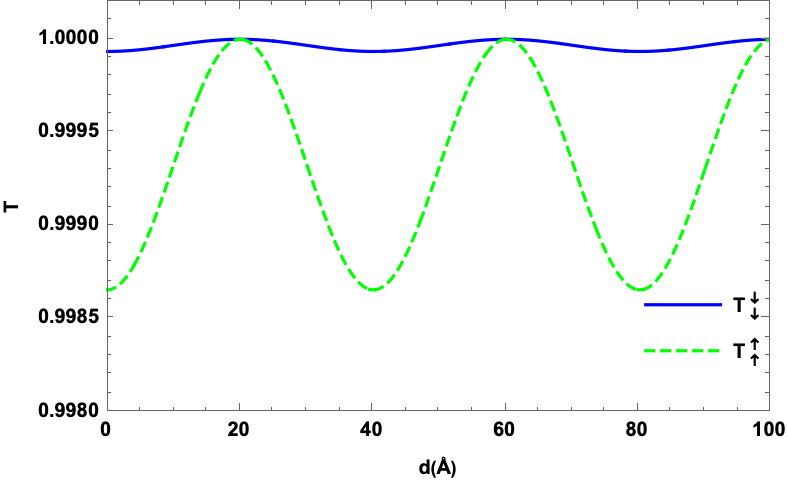}
				\label{subfigure5e}}{
				\subfloat[]{
					\includegraphics[scale=0.42]{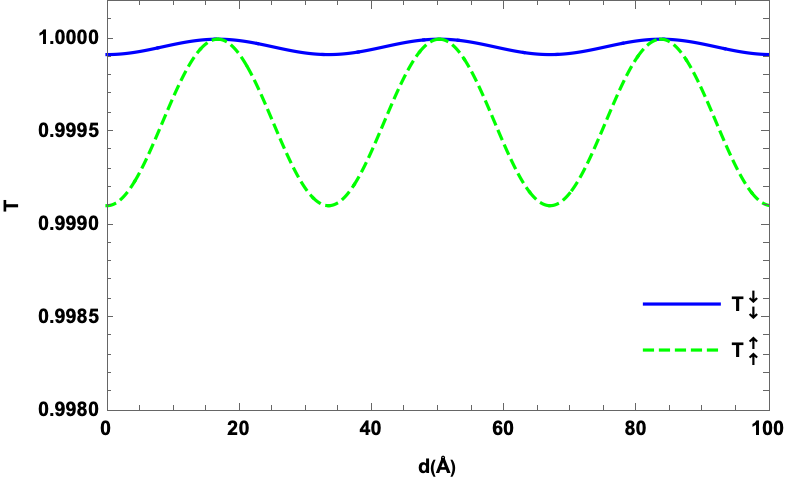}
					\label{subfigure5f}}}}}
	\caption{(color online) Transmission $T_{\uparrow}^{\uparrow}$ (green) and $T_{\downarrow}^{\downarrow}$ (blue) 
		in terms of the distance $d$ for 
		$\phi=\pi$,  $r_{0}=10\ \text{\AA{}}$, $N = 1$ with
		different values of energy    (a,d): $E=\pm 0.32$ eV,  (b,e): $E=\pm 0.5$ eV, (c,f): $E=\pm 0.6$ eV.}
	\label{fig5}
\end{figure}

We plot the two transmission channels (spin up and spin down) in terms of $d$ for positive and negative energy values with $N=1$ in Fig. \ref{fig5}.   In Fig.~\ref{subfigure5a} with $E=0.32$ eV, the transmission with spin up is dominant compared to the transmission with spin down and takes the value 1. On the other hand, the transmission with spin down shows  sinusoidal form with two maximum  and minimum values by varying around the value 1. By increasing the energy, Figs.~(\ref{subfigure5b}, \ref{subfigure5c}) show the increase in the number of oscillations as well as the number of maximum and minimum for the same interval of significance. As a result, there is a small difference in the amplitude of transmission channels for a single cell, but its behavior varies with distance $d$, proving \eqref{eq56}. In Figs.~(\ref{subfigure5d}, \ref{subfigure5e}, \ref{subfigure5f}) show that a single superlattice cell acts as a transparent selective permeable. Indeed, it allows passing the flux of electrons with spin up in one direction and the flux of electrons with spin down in the opposite direction.

\begin{figure}[H]
	\centerline{
		\subfloat[]{
			\includegraphics[scale=0.42]{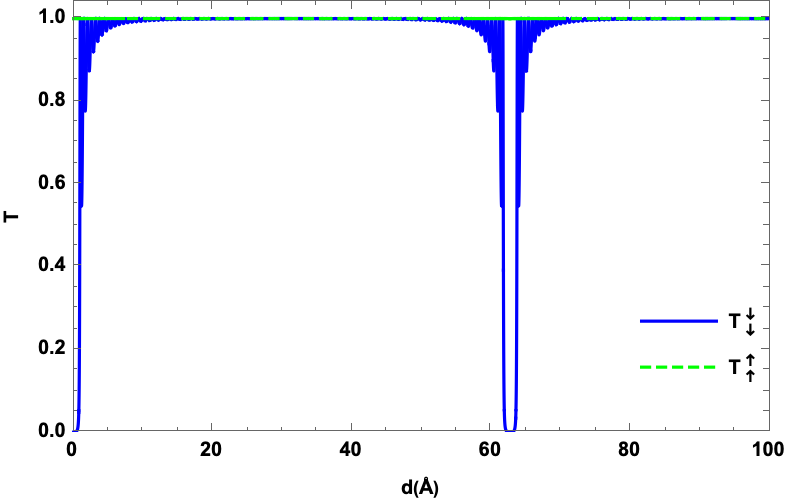}
			\label{subfigure6a}}
		{\subfloat[]{
				\includegraphics[scale=0.42]{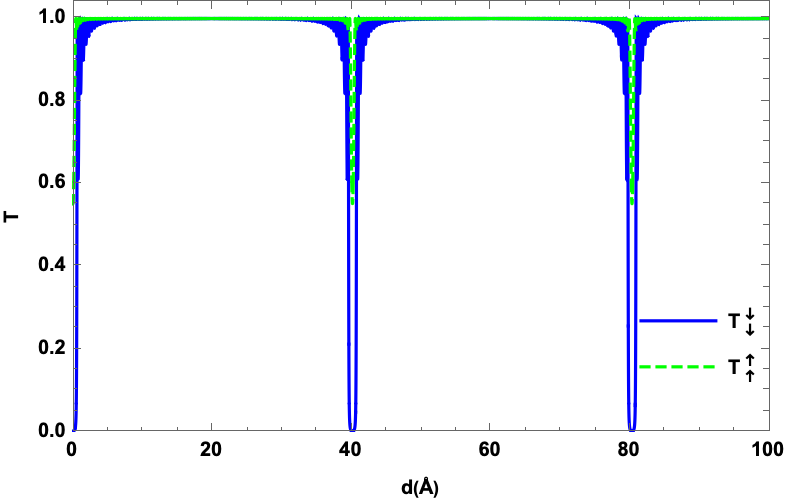}
				\label{subfigure6b}}
			{	\subfloat[]{
					\includegraphics[scale=0.42]{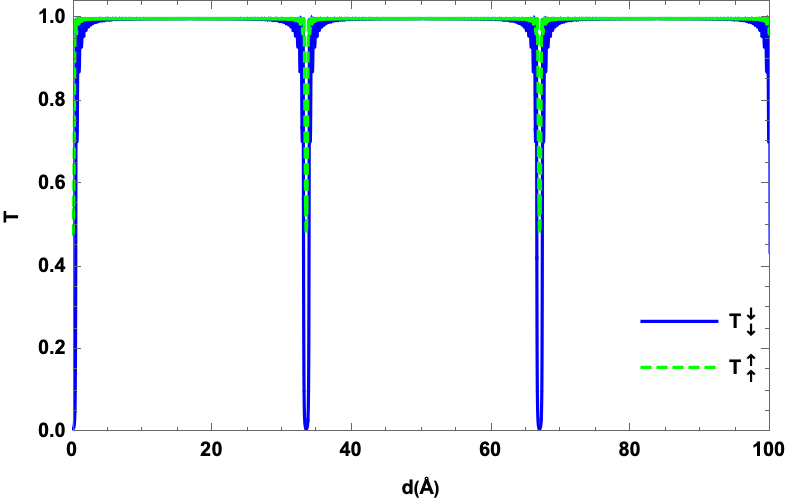}
					\label{subfigure6c}}}}}
	\centerline{
		\subfloat[]{
			\includegraphics[scale=0.42]{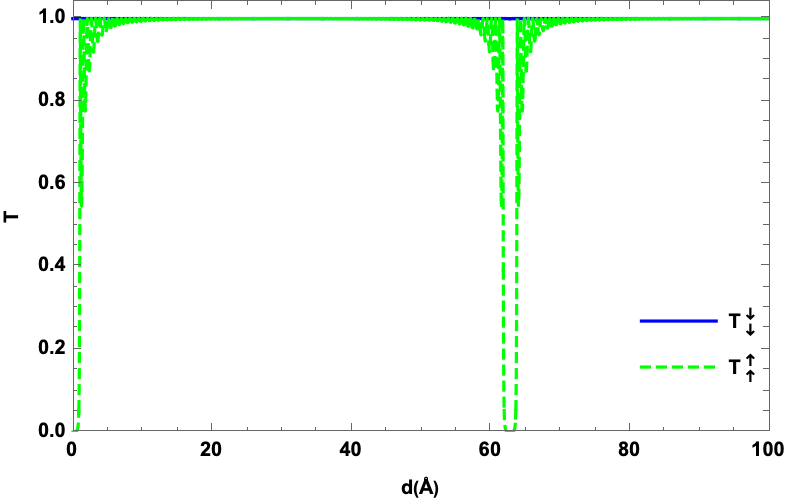}
			\label{subfigure6d}}{
			\subfloat[]{
				\includegraphics[scale=0.42]{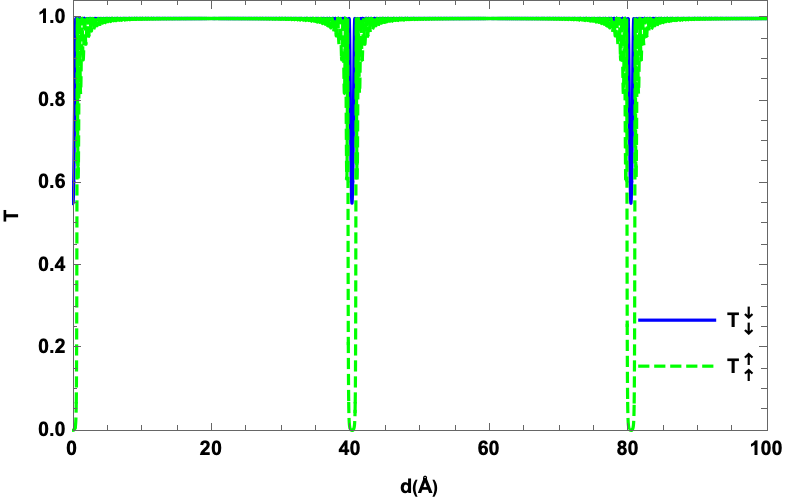}
				\label{subfigure6e}}{
				\subfloat[]{
					\includegraphics[scale=0.42]{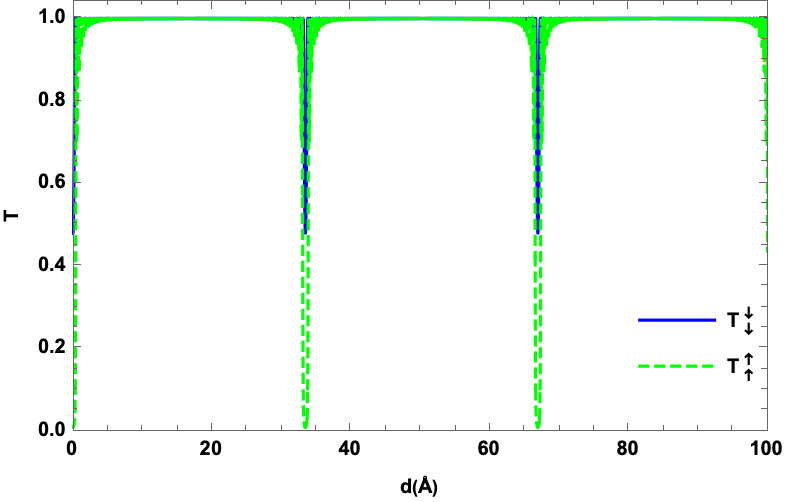}
					\label{subfigure6f}}}}}
	\caption{(color online) 
		The same as in Fig. \ref{fig5}, with
		$N = 100$.}
	\label{fig6}
\end{figure}

 In Fig.~\ref{fig6}, we consider the same situation as in Fig.~\ref{fig5} but with $N=100$ cells. Indeed, for $E=\pm 0.32$ eV, Figs. (\ref{subfigure6a}, \ref{subfigure6d}) show the appearance of two peaks that reach zero value, one at $d=0$ and the second at $d=63\ \text{\AA{}}$, corresponding to a minimal transmission for a single cell in Figs.~(\ref{subfigure5a}, \ref{subfigure5d}). 
In the case of 
 $E=\pm 0.5$ eV in Figs.~(\ref{subfigure6b}, \ref{subfigure6e}), the peaks shift back, resulting in other peaks that are completely canceled where the transmission for one cell is minimal as shown in  Figs. ~(\ref{subfigure5b}, \ref{subfigure5e}). 
 Figs.~(\ref{subfigure6c}, \ref{subfigure6f}) show an increase in peaks when the energy is increased to $E =\pm 0.6$ eV. As a result, the choice of energy has an effect on the behavior of the transmission channels. 
 More importantly,  the difference between two successive peaks allows us to determine the corresponding distance $d$ at which the transmission is suppressed. Indeed, \eqref{eq56} gives us the period  with the transmission being null. This is
 	\begin{align}
 		\Delta d= d_{l+1}-d_{l}= \frac{\gamma \pi}{E}
 	\end{align}
because $\gamma=6.39\ \angstrom\ \electronvolt$, it is obvious that $E \Delta d $= cons. As a result, we can exactly find the points at which the transmissions are suppressed for the three values of energy chosen. Indeed, for $E=\pm 0.32, 0.5, 06$ eV, we have $\Delta d=63, 40.12, 33.4 \ \text{\AA{}}$ . This tells us that distance $d$ can be used as a control to adjust the transmissions.

 	\begin{figure}[H]
 		\centerline{
 			\subfloat[]{
 				\includegraphics[scale=0.4]{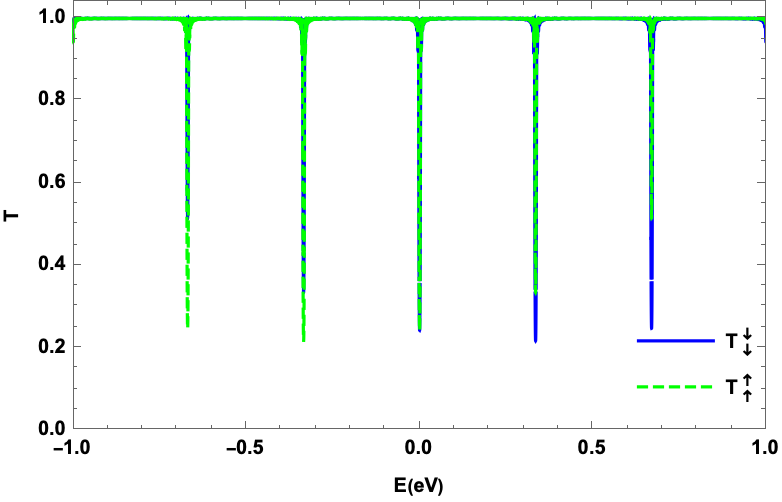}
 				\label{subfigurea4}}
 			{\subfloat[]{
 					\includegraphics[scale=0.4]{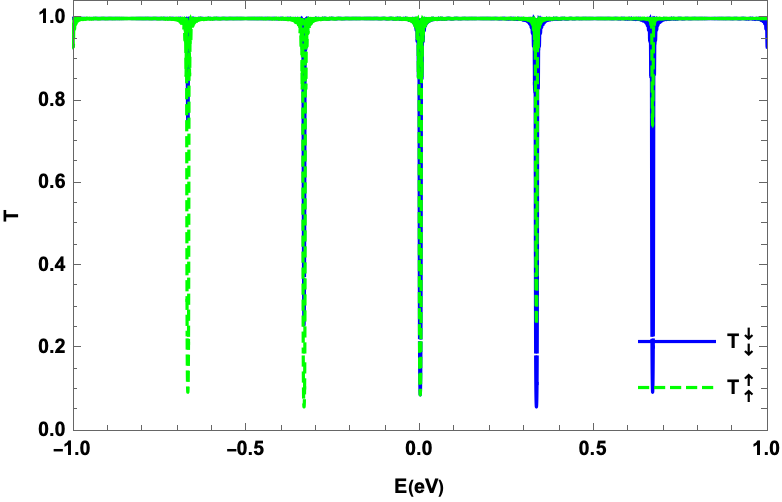}
 					\label{subfigureb4}}
 				{	\subfloat[]{
 						\includegraphics[scale=0.4]{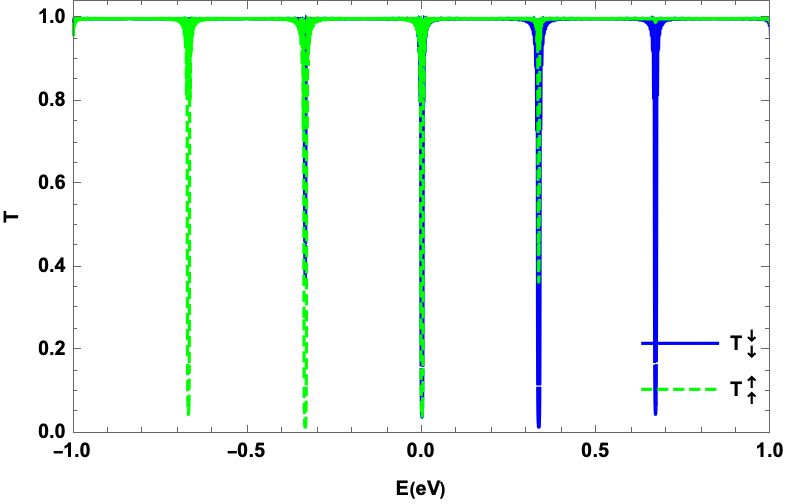}
 						\label{subfigurec4}}}
 				
 		}}
 		\caption{(color online) 
 		 Transmission channels $T_{\uparrow}^{\uparrow}$ (green) and $T_{\downarrow}^{\downarrow}$ (blue) 
 		versus incident energy $E$,  (a): $\phi=\pi/3$, (b): $\phi=\pi/2$, (c): $\phi=2\pi/3$, with $r_0=10\ \text{\AA{}} $, $d=60\ \text{\AA{}}$ and $N=100$.}
 		\label{figa4r}
 	\end{figure}
 	
To demonstrate the effect of angle $\phi$ on the behavior of the transmission, three values of $\phi$ are taken into account in Fig.~\ref{figa4r}. Indeed, in Fig.~\ref{subfigurea4} with $\phi=\pi/3$, we show that transmission suppression is not quite perfect even if the cell number is equal $N=100$. When $\phi$ is increased to  $\pi/2$ and $2\pi/3$, the transmission approaches $0$, as shown in Figs.~\ref{subfigureb4} and \ref{subfigurec4}, respectively. Just to recall, when $\phi = \pi$, the transmission is completely suppressed, as shown in Fig.~\ref{fig6}.

 	\begin{figure}[H]
 		\centerline{
 			\subfloat[]{
 				\includegraphics[scale=0.4]{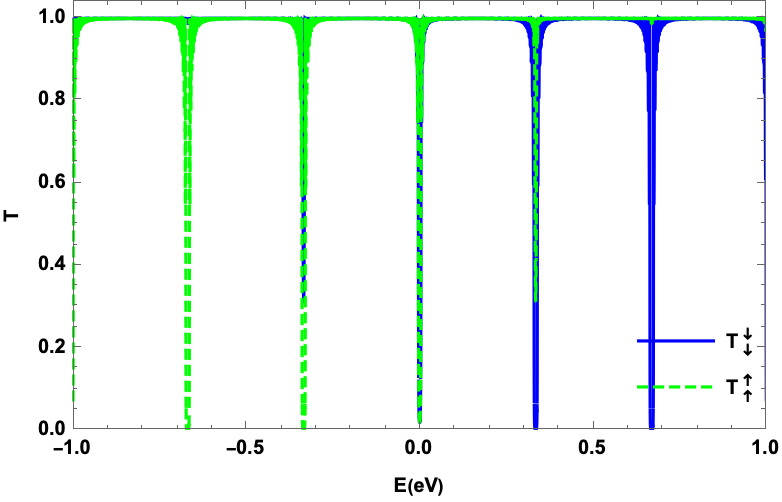}
 				\label{subfigurea5}}
 			{\subfloat[]{
 					\includegraphics[scale=0.4]{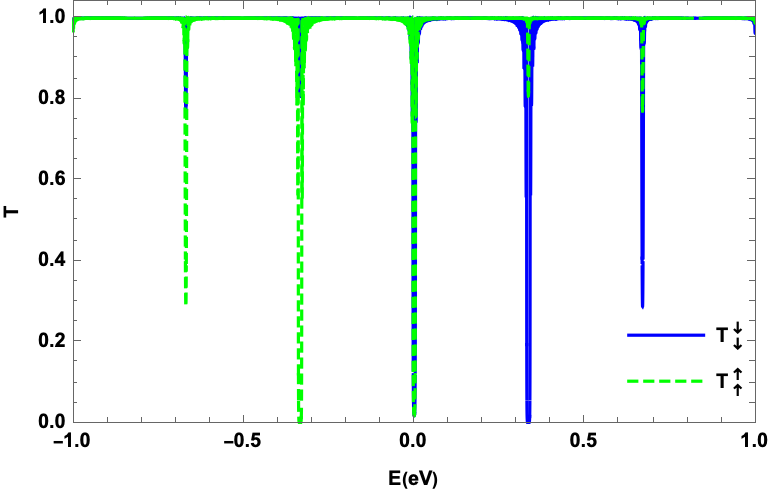}
 					\label{subfigureb5}}
 				{	\subfloat[]{
 						\includegraphics[scale=0.4]{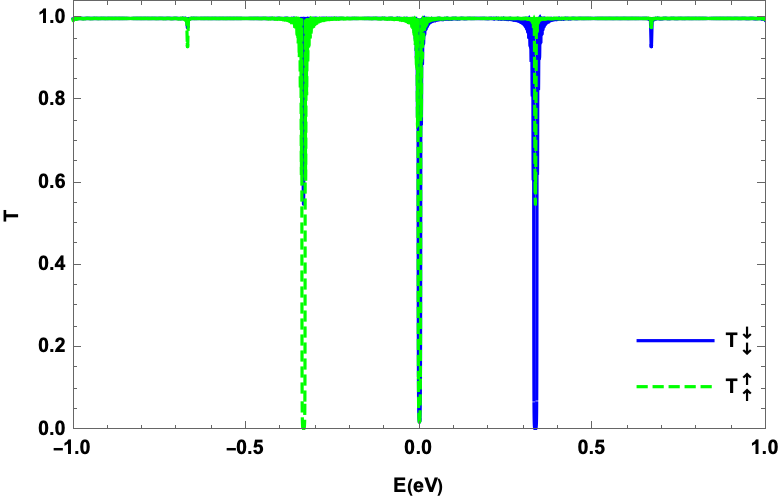}
 						\label{subfigurec5}}}
 				
 		}}
 		\caption{(color online) 
 			 Transmission channels $T_{\uparrow}^{\uparrow}$ (green) and $T_{\downarrow}^{\downarrow}$ (blue) 
 			versus incident energy $E$,  (a): $r_0=5\ \text{\AA{}} $, (b): $r_0=12\ \text{\AA{}} $, (c): $r_0=15\ \text{\AA{}} $, with $\phi=\pi$, $d=60\ \text{\AA{}}$ and $N=100$.}
 		\label{figa5r}
 	\end{figure}	
Another parameter that can modify the behavior of the transmission is the radius of the curvature. In this case, Fig. \ref{figa5r} depicts the transmission's behavior for three different values of the radius $r_0$. In Fig.~\ref{subfigurea5}, with $r_0=5\ \text{\AA{}} $ we show that the transmission vanishes completely over a large energy range. The transmission suppression interval is reduced by the appearance of peaks that do not reach the value $0$ for $r_0=12\   \text{\AA{}} $ and $r_0=15\   \text{\AA{}} $, as shown in 
  	Figs.~\ref{subfigureb5} and  \ref{subfigurec5}, respectively. We found that the smaller the ripple radius, the larger the filtering effect  at $N =100$.

	\section*{Conclusion}\label{sec5}
	
We carefully examined the two electron transmission channels (spin up and spin down) through the superlattice created on a regular basis from wavy graphene with normal incidence ($k_y = 0$). By applying the boundary continuity between the different regions of a single cell, we have deduced the transfer matrix of our cell and, subsequently, that of the superlattice. Then, we calculated and analyzed transmissions with spin-up and spin-down in the presence of system parameters. Indeed, we have demonstrated that the introduction of the distance $d$ has a remarkable effect on the propagation and scattering of electrons with different polarizations. With a single cell, the variation in amplitude is small, but the behavior of the transmission changes with the creation of oscillations which attenuate according to energy.

With increasing cell number and distance $ d $, the effect becomes significant. Furthermore, our numerical results show that the transmission with the spin down is strongly suppressed in comparison to the transmission with the spin up for values $E > 0 $, but the transmission with the spin up is strongly suppressed in comparison to the transmission with the spin down for $E <0 $. On the other hand, in the case where $d$ is different from $0$, the transmission of electrons with spin down is suppressed almost the same as with spin up but with a higher number of peaks. The interval between peaks is getting narrower and peaks also appear where transmission is minimal, in the case of a single cell. The distance $d$ is the responsible parameter which controls the step between these peaks. On the other hand, the number of cells controls the intensity of transmission suppression.

The number of oscillations increases as $d$ increases, resulting in the appearance of several minima, which are canceled out when the number of cells $N$ is large enough. This result has not been observed previously in the literature \cite{pudlak2015cooperative}. The greater the number of cells (we assume that the ripples are identical), the stronger the suppression effect. Thus, the distance $d$ between the ripples (concave and convex) allows us to control the degree of suppression of transmission with spin up/down in a range of energy.

The present work will not remain at this stage. Indeed, as it becomes known that  the strain applied to graphene can generate the Landau levels in a similar way to an applied magnetic field \cite{Settnes2019, Castro2017}, one can ask how the strain can modify the results obtained here. The answer to this question is under investigation.

	\section*{Refernces}

	\end{document}